\documentstyle[aps]{revtex}

\begin{document}
\title{Quantization of The Electroweak Theory in The Hamiltonian Path-Integral
Formalism}
\author{Jun-Chen Su}
\address{Center for Theoretical Physics, Department of Physics, Jilin\\
University,Changchun 130023, People's Republic of China}
\maketitle

\begin{abstract}
The quantization of the SU(2)$\times $U(1) gauge-symmetric electroweak
theory is performed in the Hamiltonian path-integral formalism. In this
quantization, we start from the Lagrangian given in the unitary gauge in
which the unphysical Goldstone fields are absent, but the unphysical
longitudinal components of the gauge fields still exist. In order to
eliminate the longitudinal components, it is necessary to introduce the
Lorentz gauge conditions as constraints. These constraints may be
incorporated into the Lagrangian by the Lagrange undetermined multiplier
method. In this way, it is found that every component of a four-dimensional
vector potential has a conjugate counterpart. Thus, a Lorentz-covariant
quantization in the Hamiltonian path-integral formalism can be well
accomplished and leads to a result which is the same as given by the
Faddeev-Popov approach of quantization.

\begin{description}
\item[PACS:]  11.15.-qy; 11.10.Gh; 12.20.-m
\end{description}
\end{abstract}

In our preceding paper$^{[1]}$, the quantum electroweak theory without
involving the Goldstone bosons$^{[2,3]}$ was established starting from the
Lagrangian given in the unitary gauge by the Faddeev-Popov approach$^{[4]}$
in the Lagrangian path-integral formalism. The quantum theory given in the $%
\alpha $-gauge shows good renormalizability. The unitarity is ensured by the
limiting procedure: $\alpha \rightarrow \infty $ by which the results
calculated in the $\alpha $-gauge will be converted to the physical ones as
should be obtained in the unitary gauge. An important feature of the theory
is that the theory established is still of SU(2)$\times $U(1) gauge
symmetry. To check this point, in this paper, the quantization of the theory
will be carried out in the Hamiltonian path-integral formalism$^{[5-7]}$.
This quantization, unlike the quantization by the Faddeev-Popov approach,
does not concern the gauge transformation and the gauge-invariance of the
Lagrangian chosen for the quantization.

For simplicity, we limit ourself to discuss the electroweak interaction
system for one generation of leptons. the Lagrangian is$^{[2,3,8]}$ 
\begin{equation}
{\cal L}={\cal L}_g+{\cal L}_f+{\cal L}_\phi  \eqnum{1}
\end{equation}
where 
\begin{equation}
{\cal L}_g=-\frac 14F^{\alpha \mu \nu }F_{\mu \nu }^\alpha  \eqnum{2}
\end{equation}
\begin{equation}
{\cal L}_f=\overline{L}i\gamma ^\mu D_\mu L+\overline{l}_Ri\gamma ^\mu D_\mu
l_R  \eqnum{3}
\end{equation}
and 
\begin{equation}
{\cal L}_\phi =\frac 12(D^\mu \phi )^{+}(D_\mu \phi )-\frac 12\mu ^2\phi
^{+}\phi -\frac 14\lambda (\phi ^{+}\phi )^2-\frac{f_l}{\sqrt{2}}(\overline{L%
}\phi l_R+\overline{l}_R\phi ^{+}L)  \eqnum{4}
\end{equation}
are respectively the parts of the Lagrangian for the gauge boson, lepton and
scalar particle fields. In the above, 
\begin{equation}
F_{\mu \nu }^\alpha =\partial _\mu A_\nu ^\alpha -\partial _\nu A_\mu
^\alpha +g\varepsilon ^{\alpha \beta \gamma }A_\mu ^\beta A_\nu ^\gamma 
\eqnum{5}
\end{equation}
where $\alpha =0,1,2,3,$ A$_\mu ^i=$ (A$_\mu ^a,B_\mu ),$A$_\mu ^a$ and $%
B_\mu $ are the SU(2) and U(1) gauge fields respectively and the Levi-Civita
tensor is defined by 
\begin{equation}
\varepsilon ^{\alpha \beta \gamma }=\{ 
\begin{array}{c}
\epsilon ^{abc}\text{ , if }\alpha ,\beta ,\gamma =a,b,c=1,2\text{ or }3; \\ 
0\text{ , if }\alpha ,\beta \text{ and/or }\gamma =0
\end{array}
\eqnum{6}
\end{equation}
\begin{equation}
L=\left( 
\begin{array}{c}
\nu _L \\ 
l_L
\end{array}
\right)  \eqnum{7}
\end{equation}
here $\nu _L$ represents the left handed neutrino field and $l_L$ the
left-handed lepton field, $l_R$ is the right-handed lepton field , 
\begin{equation}
D_\mu =\partial _\mu -ig\frac{\tau ^a}2A_\mu ^a-ig^{\prime }\frac Y2B_\mu 
\eqnum{8}
\end{equation}
is the covariant derivative in which $\frac{\tau ^a}2$ and $\frac Y2$ are
the generators of SU(2)$_T$ and U(1)$_Y$ gauge groups, respectively, and 
\begin{equation}
\phi (x)=\left( 
\begin{array}{c}
0 \\ 
H(x)+v
\end{array}
\right)  \eqnum{9}
\end{equation}
is the scalar field doublet, $g,g^{\prime },\mu ^2,\lambda $ and $f_l$ are
the coupling constants with a relation $v=\sqrt{-\mu ^2/\lambda }$.

In above Lagrangian, still exist the unphysical longitudinal parts of the
gauge fields. They may completely be eliminated by introducing the Lorentz
conditions 
\begin{equation}
\chi ^\alpha \equiv \partial ^\mu A_\mu ^\alpha =0  \eqnum{10}
\end{equation}

According to the general procedure of dealing with constrained systems $%
^{[5,6]}$, the Lorentz conditions, as constraints, may be incorporated into
the Lagrangian by the Lagrange multiplier method 
\begin{equation}
{\cal L}_\lambda ={\cal L}+\lambda ^\alpha \partial ^\mu A_\mu ^\alpha 
\eqnum{11}
\end{equation}
where ${\cal L}$ is the Lagrangian shown in Eqs.(1)-(4) and $\lambda ^\alpha 
$ are the Lagrange multipliers. The above Lagrangian will suitably be chosen
to be the starting point of performing the Hamiltonian path-integral
quantization. The advantage of the Lagrangian is that it provides each
component of a vector potential a canonically conjugate counterpart. In
fact, according to the usual definition of the canonical momentum (density)
of a field variable, it is found 
\begin{equation}
\Pi _\mu ^\alpha =\frac{\partial {\cal L}_\lambda }{\partial \stackrel{%
\bullet }{A}^{\alpha \mu }}=F_{\mu 0}^\alpha +\lambda ^\alpha \delta _{\mu
0}=\{ 
\begin{array}{c}
F_{k0}^\alpha \equiv E_k^\alpha \text{ , if }\mu =k=1,2,3 \\ 
\lambda ^\alpha \equiv -E_0^\alpha \text{ , if }\mu =0
\end{array}
\eqnum{12}
\end{equation}
here, as we see, the Lagrange multipliers act as the time-components of the
conjugate momenta. With the conjugate momenta defined above and the
conjugate momenta for fermion fields and the Higgs field which are defined
as 
\begin{eqnarray}
\Pi _l &=&\frac{\partial {\cal L}_\lambda }{\delta \stackrel{\bullet }{l}}%
=il^{+}\text{ , }\overline{\Pi }_l=\frac{\partial {\cal L}_\lambda }{\delta 
\stackrel{\bullet }{\stackrel{\_}{l}}}=0  \nonumber \\
\Pi _\nu &=&\frac{\partial {\cal L}_\lambda }{\delta \stackrel{\bullet }{\nu 
}}=i\nu _L^{+}\text{ , }\overline{\Pi }_\nu =\frac{\partial {\cal L}_\lambda 
}{\delta \stackrel{\bullet }{\stackrel{\_}{\nu }}}=0  \eqnum{13}
\end{eqnarray}
and 
\begin{equation}
\Pi _H=\frac{\partial {\cal L}_\lambda }{\delta \stackrel{\bullet }{H}}=%
\stackrel{\bullet }{H}  \eqnum{14}
\end{equation}
according to the standard procedure, the Lagrangian in Eq.(11) may readily
be recast in the first-order form 
\begin{equation}
{\cal L}_\lambda =E^{\alpha \mu }\stackrel{\bullet }{A}_\mu ^\alpha +\Pi _l%
\stackrel{\bullet }{l}+\Pi _\nu \stackrel{\bullet }{\nu }+\Pi _H\stackrel{%
\bullet }{H}-{\cal H}+A_0^\alpha \varphi ^\alpha -E_0^\alpha \chi ^\alpha 
\eqnum{15}
\end{equation}
where 
\begin{equation}
{\cal H}={\cal H}_g+{\cal H}_f+{\cal H}_H  \eqnum{16}
\end{equation}
is the Hamiltonian (density) in which 
\begin{equation}
{\cal H}_g=\frac 12[(E_k^\alpha )^2+(B_k^\alpha )^2]  \eqnum{17}
\end{equation}
here 
\begin{equation}
B_k^\alpha =-\frac 12\varepsilon _{ijk}F_{ij}^\alpha  \eqnum{18}
\end{equation}
\begin{equation}
{\cal H}_f=-i\overline{\nu }_L\gamma ^k\partial _k\nu _L-i\overline{l}\gamma
^k\partial _kl-\frac 12\overline{L}\gamma ^k(g\tau ^aA_k^a-g^{\prime
}B_k)L+g^{\prime }\overline{l}_R\gamma ^kB_kl_R  \eqnum{19}
\end{equation}
and 
\begin{eqnarray}
{\cal H}_H &=&\frac 12\Pi _H^2+\frac 12\left( \nabla H\right) ^2+\frac 12\mu
^2\left( H+v\right) ^2+\frac \lambda 4\left( H+v\right) ^4  \nonumber \\
&&\ \ \ \ \ \ +\frac 14\{g^2[(A_k^1)^2+(A_k^2)^2]+(g^{\prime
}B_k-gA_k^3)^2\}\left( H+v\right) ^2  \eqnum{20} \\
&&\ \ \ \ \ \ +\frac{f_l}{\sqrt{2}}\overline{l}l\left( H+v\right)  \nonumber
\end{eqnarray}
\begin{equation}
\varphi ^\alpha =\varphi _g^\alpha +\varphi _f^\alpha +\varphi _H^\alpha 
\eqnum{21}
\end{equation}
in which 
\begin{equation}
\varphi _g^\alpha =\partial ^\mu E_\mu ^\alpha -g\varepsilon ^{\alpha \beta
\gamma }E_k^\beta A^{\gamma k}  \eqnum{22}
\end{equation}
\begin{equation}
\varphi _f^\alpha =\{ 
\begin{array}{c}
\frac 12g\overline{L}\gamma _0\tau ^aL\text{ , if }\alpha =a=1,2,3; \\ 
-\frac 12g^{\prime }(\overline{L}\gamma _0L+2\overline{l}_R\gamma _0l_R)%
\text{ , if }\alpha =0
\end{array}
\eqnum{23}
\end{equation}
and 
\begin{equation}
\varphi _H^\alpha =\{ 
\begin{array}{c}
\frac 14g(gA_0^a-\delta ^{a3}g^{\prime }B_0)\left( H+v\right) ^2\text{ , if }%
\alpha =a=1,2,3; \\ 
\frac 14g^{\prime }(g^{\prime }B_0-gA_0^3)\left( H+v\right) ^2\text{ , if }%
\alpha =0
\end{array}
\eqnum{24}
\end{equation}
and $\chi ^\alpha $ was defined in Eq.(10).

From the structure of the Lagrangian in Eq.(15), it is clearly seen that the
last two terms in Eq.(15) are actually given by incorporating the constraint
conditions 
\begin{equation}
\varphi ^\alpha =0  \eqnum{25}
\end{equation}
and that in Eq.(10) into the Lagrangian by the Lagrange multiplier method.
These constraint conditions may be derived from the stationary condition of
the action $S_\lambda =\int d^4x{\cal L}_\lambda (x)$. From this derivation,
one may also obtain equations of motion (see Appendix A) in which there are
time-derivatives of the dynamical field variables $A_k^a,B_k,F_{k0}^a=E_k^a$
and $B_{k0}=E_k^0$; whereas in Eqs.(10) and (25) there are no such
derivatives. Therefore, Eqs.(10) and (25) can only be identified with the
constraint equations. Since $\partial ^\mu E_\mu ^\alpha $ =$\partial ^\mu
E_{L\mu }^\alpha $ and $\partial ^\mu A_\mu ^\alpha =\partial ^\mu A_{L\mu
}^\alpha $ where $E_{L\mu }^\alpha $ and $A_{L\mu }^\alpha $ are the
longitudinal parts of the canonical variables $E_\mu ^\alpha $ and $A_\mu
^\alpha $ respectively, we see, the conditions in Eq.(26) and (27) are
responsible respectively for constraining the unphysical longitudinal parts
of the canonical variables $E_{L\mu }^\alpha $ and $A_{L\mu }^\alpha $ .
Therefore, only the transverse parts of the variables, $E_{T\mu }^\alpha $
and $A_{T\mu }^\alpha ,$ can be viewed as independent dynamical field
variables. Because each of the transverse vectors $E_{T\mu }^\alpha $ and $%
A_{T\mu }^\alpha $ contains three independent components, they are
sufficient to describe the polarization states of the massive gauge fields.

Let us turn to the solutions of the equations (10) and (25). The solution of
equation (10), as is well-known, is 
\begin{equation}
A_{L\mu }^\alpha =0  \eqnum{26}
\end{equation}
For the equation (25) with the $\varphi ^\alpha $ being represented in
Eqs.(21)-(24), we would like to note that the function $\varphi _g^\alpha $
in Eq.(22) can also be written in the form of Lorentz-covariance 
\begin{equation}
\varphi _g^\alpha =\partial ^\mu E_\mu ^\alpha -g\varepsilon ^{\alpha \beta
\gamma }E_\mu ^\beta A^{\gamma \mu }  \eqnum{27}
\end{equation}
This is because the added term $g\varepsilon ^{\alpha \beta \gamma
}E_0^\beta A_0^\gamma $ in the above gives a vanishing contribution to the
Lagrangian (see the term $A_0^\alpha \varphi ^\alpha $ in Eq.(15)) due to
the identity $\varepsilon ^{\alpha \beta \gamma }A_0^\alpha A_0^\gamma
\equiv 0$. On substituting Eq.(26) into Eq.(25) and noticing that the
longitudinal vector $E_{L\mu }^\alpha $ can always be represented as 
\begin{equation}
E_{L\mu }^\alpha =\partial _\mu Q^\alpha  \eqnum{28}
\end{equation}
where $Q^a$ is a scalar function, the equation in Eq.(25) may be written in
the form$^{[4,5]}$ 
\begin{equation}
K^{\alpha \beta }(x)Q^\beta (x)=R^\alpha (x)  \eqnum{29}
\end{equation}
where 
\begin{equation}
K^{\alpha \beta }=\delta ^{\alpha \beta }\Box -g\varepsilon ^{\alpha \beta
\gamma }A_T^{\gamma \mu }\partial _\mu  \eqnum{30}
\end{equation}
and 
\begin{equation}
R^\alpha =g\varepsilon ^{\alpha \beta \gamma }E_{T\mu }^\beta A_T^{\gamma
\mu }-\varphi _f^\alpha (A_{0T}^\alpha )-\varphi _H^\alpha (A_{0T}^\alpha ) 
\eqnum{31}
\end{equation}
here $\varphi _f^\alpha (A_{0T}^\alpha )$ and $\varphi _H^\alpha
(A_{oT}^\alpha )$ are defined in Eqs.(23) and (24) with the $A_0^\alpha $
being replaced by $A_{0T}^\alpha $. With the aid of the Green's function $%
\triangle ^{\alpha \beta }(x-y)$ (the ghost particle propagator) which
satisfies the equation 
\begin{equation}
K^{\alpha \gamma }(x)\triangle ^{\gamma \beta }(x-y)=\delta ^{\alpha \beta
}\delta ^4(x-y)  \eqnum{32}
\end{equation}
the solution of equation (29) is found to be 
\begin{equation}
Q^\alpha (x)=\int d^4y\triangle ^{\alpha \beta }(x-y)R^\beta (y)  \eqnum{33}
\end{equation}
Inserting this result into Eq.(28), we get 
\begin{equation}
E_{L\mu }^\alpha =E_{L\mu }^\alpha (A_{T\mu }^\alpha ,E_{T\mu }^\alpha
,\cdots )  \eqnum{34}
\end{equation}
which is a functional of the independent field variables $A_{T\mu }^\alpha
,E_{T\mu }^\alpha $ and others.

Now, we are ready to carry out the quantization in the Hamiltonian
path-integral formalism. In accordance with the basic idea of path-integral
quantization$^{[5-7]}$, we are allowed to directly write out an exact
generating functional of Green's functions by making use of the Hamiltonian
which is expressed in terms of the independent field variables 
\begin{equation}
Z\left[ J\right] =\frac 1N\int D\left( \Pi ^{*},\Phi ^{*}\right) \exp
\{i\int d^4x(\Pi ^{*}\stackrel{}{\cdot \stackrel{\bullet }{\Phi ^{*}}}-{\cal %
H}^{*}\left( \Pi ^{*},\Phi ^{*}\right) +J^{*}\cdot \Phi ^{*})\}  \eqnum{35}
\end{equation}
where $\Phi ^{*}$ and $\Pi ^{*}$stand for all the independent variables $%
(A_{T\mu }^\alpha ,\nu ,l,H)$ and the conjugate ones $(E_{T\mu }^\alpha ,\Pi
_\nu ,\Pi _l,\Pi _H)$ respectively, $J$ denotes the external sources and $%
{\cal H}^{*}(\Pi ^{*},\Phi ^{*})$ is the Hamiltonian defined by$^{[6]}$ 
\begin{equation}
{\cal H}^{*}(\Pi ^{*},\Phi ^{*})={\cal H}(\Pi ,\Phi )|_{A_{L\mu }^\alpha
=0,E_{L\mu }^\alpha =E_{L\mu }^\alpha (\Pi ^{*},\Phi ^{*})}  \eqnum{36}
\end{equation}
This Hamiltonian, as it stands, has a complicated functional structure which
is not convenient for establishing the perturbation theory. Therefore, one
still expects to represent the generating functional through the full
vectors $A_\mu ^\alpha $ and $E_\mu ^\alpha $. For this purpose, it is
necessary to introduce the delta-functionals into the generating functional
like this$^{[6]}$ 
\begin{eqnarray}
Z\left[ J\right] &=&\frac 1N\int D\left( \Pi ,\Phi \right) \delta \left[
A_L\right] \delta [E_L-E_L(\Pi ^{*},\Phi ^{*})]  \nonumber \\
&&\ \ \ \times \exp \{i\int d^4x[\Pi \cdot \Phi \stackrel{.}{-{\cal H}\left(
\Pi ,\Phi \right) }+J\cdot \Phi ]\}  \eqnum{37}
\end{eqnarray}
where $\Phi =(A_\mu ^\alpha ,\nu ,l,H)$ and $\Pi =(E_\mu ^\alpha ,\Pi _\nu
,\Pi _l,\Pi _H)$.

The delta-functionals in Eq.(37) can be expressed as a useful form as follows%
$^{[6]}$(see appendix B) 
\begin{equation}
\delta \left[ A_L\right] \delta \left[ E_L-E_L\left( \Pi ^{*},\Phi
^{*}\right) \right] =\det M[A]\delta [\varphi ]\delta [\chi ]  \eqnum{38}
\end{equation}
where $\delta [\varphi ]$ and $\delta [\chi ]$ represent the constraint
conditions in Eq.(10) and (25) and M[A] is a matrix whose elements are given
by the following Poisson bracket$^{[6]}$ 
\begin{eqnarray}
M^{\alpha \beta }(x,y) &=&\{\varphi ^\alpha (x),\chi ^\beta (y)\}  \nonumber
\\
&=&\int d^4z\{\frac{\delta \varphi ^\alpha (x)}{\delta A_\mu ^\gamma (z)}%
\frac{\delta \chi ^\beta (y)}{\delta E^{\gamma \mu }(z)}-\frac{\delta
\varphi ^\alpha (x)}{\delta E_\mu ^\gamma (z)}\frac{\delta \chi ^\beta (y)}{%
\delta A^{\gamma \mu }(z)}\}  \eqnum{39}
\end{eqnarray}
These matrix elements are easily evaluated by using the expressions denoted
in Eqs.(10), (21), (27), (23) and (24). The result is 
\begin{equation}
M^{\alpha \beta }(x,y)=\partial _x^\mu \left[ D_\mu ^{\alpha \beta
}(x)\delta ^4(x-y)\right]  \eqnum{40}
\end{equation}
where 
\begin{equation}
D_\mu ^{\alpha \beta }(x)=\delta ^{\alpha \beta }\partial _\mu
^x-g\varepsilon ^{\alpha \beta \gamma }A_{\mu ^{}}^\gamma (x)  \eqnum{41}
\end{equation}

Upon inserting the relation in Eq.(38) into Eq.(37) and employing the
Fourier representation for $\delta [\varphi ]$%
\begin{equation}
\delta [\varphi ]=\int D(\frac \eta {2\pi })e^{i\int d^4x\eta ^\alpha
(x)\varphi ^\alpha (x)}  \eqnum{42}
\end{equation}
we have 
\begin{eqnarray}
Z\left[ J\right] &=&\frac 1N\int D\left( \Pi ,\Phi \right) D(\frac{d\eta }{%
2\pi })\det M[A]\delta [\chi ]  \nonumber \\
&&\ \ \ \ \times \exp \{i\int d^4x(\Pi \cdot \stackrel{}{\stackrel{\bullet }{%
\Phi }+\eta ^\alpha }\varphi ^\alpha -{\cal H}+J\cdot \Phi )\}  \eqnum{43}
\end{eqnarray}
For later convenience, the $E_0^\alpha $-dependent terms will be extracted
from the first two terms in the above exponent and thus Eq.(45) will be
rewritten as 
\begin{eqnarray}
Z\left[ J\right] &=&\frac 1N\int D(\Pi ^{\prime },\Phi ^{\prime })D\left(
E_0,A_0\right) D(\frac \eta {2\pi })\det M\left[ A\right] \delta [\chi ] 
\nonumber \\
&&\ \ \ \ \times \exp \{i\int d^4x[E_0^\alpha (\stackrel{\bullet }{A}%
_0^\alpha -\stackrel{\bullet }{\eta }^\alpha )+\Pi ^{\prime }\cdot \stackrel{%
\bullet }{\Phi ^{\prime }}+\eta ^\alpha \overline{\varphi }^\alpha -{\cal H}%
+J\cdot \Phi ]\}  \eqnum{44}
\end{eqnarray}
where 
\begin{equation}
\Pi ^{\prime }\cdot \stackrel{\bullet }{\Phi ^{\prime }}=E^{\alpha k}%
\stackrel{\bullet }{A}_k^\alpha +\Pi _\nu \stackrel{\bullet }{\nu }+\Pi _l%
\stackrel{\bullet }{l}+\Pi _H\stackrel{\bullet }{H}  \eqnum{45}
\end{equation}
and 
\begin{equation}
\eta ^\alpha \stackrel{}{\overline{\varphi }^\alpha }=\eta ^\alpha (\partial
^kE_k^\alpha +g\varepsilon ^{\alpha \beta \gamma }A_k^\beta E^{\gamma
k}+\varphi _f^\alpha +\varphi _H^\alpha )  \eqnum{46}
\end{equation}
The integral over $E_0^\alpha $ in Eq.(44) gives the delta-functional 
\begin{equation}
\delta [\stackrel{\bullet }{A}_0^\alpha -\stackrel{\bullet }{\eta }^\alpha
]=\det \left| \partial _0\right| ^{-1}\delta [\eta ^\alpha -A_0^\alpha ] 
\eqnum{47}
\end{equation}
The determinant in the above, as a constant, may be put in the normalization
constant N. The delta-functional $\delta [\eta ^\alpha -A_0^\alpha ]$ will
be used to perform the integration over $\eta ^\alpha $ in Eq.(44). After
these manipulations, we get 
\begin{eqnarray}
Z\left[ J\right] &=&\frac 1N\int D\left( \Pi ^{\prime },\Phi ^{\prime
}\right) D\left( A_0\right) \det M[A]\delta [\chi ]  \nonumber \\
&&\ \ \ \ \times \exp \{i\int d^4x[\Pi ^{\prime }\cdot \stackrel{\bullet }{%
\Phi ^{\prime }}+A_0^\alpha \overline{\varphi }^\alpha -{\cal H}+J\cdot \Phi
]\}  \eqnum{48}
\end{eqnarray}
In the above expression, the integrals over $E_k^\alpha $ and $\Pi _H$ are
of Gaussion type and hence are easily calculated, giving 
\begin{equation}
\int D\left( E_k^\alpha \right) e^{-i\int d^4x\left[ \frac 12\left(
E_k^\alpha \right) ^2+E_k^\alpha F_{ok}^\alpha \right] }=e^{i\int d^4x\frac 1%
2\left( F_{k0}^\alpha \right) ^2}  \eqnum{49}
\end{equation}
\begin{equation}
\int D\left( \Pi _H\right) e^{-i\int d^4x\left[ \frac 12\Pi _H^2-\Pi _H%
\stackrel{\bullet }{H}\right] }=e^{i\int d^4x\frac 12\stackrel{\bullet }{H}%
^2}  \eqnum{50}
\end{equation}
For the integrals over $\Pi _\nu $ and $\Pi _l$, the integration variables $%
\Pi _\nu $ and $\Pi _l$ will be changed to $\overline{\nu }$ and $\overline{l%
\text{.}}$ The Jacobian caused by this change, as a constant, may be put in
the constant N. On substituting Eqs.(49) and (50) in Eq.(48), it is easy to
see that in the functional integral thus obtained, except for the external
source terms, the sum of the other terms in the exponent just give the
original Lagrangian shown in Eqs.(1)-(9). Thus, we obtain 
\begin{equation}
Z[J]=\frac 1N\int D\left( \Psi ^{\prime }\right) \det M\delta [\chi
]e^{i\int d^4x[{\cal L}+J\cdot \Psi ^{\prime }]}  \eqnum{51}
\end{equation}
where $\Psi ^{\prime }=\left( \overline{l},l,\overline{\nu },\nu ,A_\mu
^\alpha ,H\right) $. When making use of the familiar expression of the
determinant$^{[4]}$ 
\begin{eqnarray}
\det M &=&\int D\left( \overline{C},C\right) e^{i\int d^4xd^4y\overline{C}%
^\alpha (x)M^{\alpha \beta }(x,y)C^\beta (y)}  \nonumber \\
\ &=&\int D\left( \overline{C},C\right) e^{i\int d^4x\overline{C}^\alpha
\partial ^\mu (D_\mu ^{\alpha \beta }C^\beta )}  \eqnum{52}
\end{eqnarray}
and the Fresnel representation for the delta-functional 
\begin{equation}
\delta [\partial ^\mu A_\mu ^{}]=\lim_{\alpha \rightarrow 0}C[\alpha ]\exp
\{-\frac i{2\alpha }\int d^4x\left( \partial ^\mu A_\mu ^\alpha \right) ^2\}
\eqnum{53}
\end{equation}
where $C[\alpha ]$ is a constant, we arrive at 
\begin{equation}
Z[J]=\frac 1N\int D(\Psi )\exp \{i\int d^4x({\cal L}_{eff}+J\cdot \Psi )\} 
\eqnum{54}
\end{equation}
where 
\begin{equation}
{\cal L}_{eff}={\cal L}-\frac 1{2\alpha }(\partial ^\mu A_\mu ^\alpha )^2+%
\overline{C}^\alpha \partial ^\mu (D_\mu ^{\alpha \beta }C^\beta ) 
\eqnum{55}
\end{equation}
is the effective Lagrangian and the limit $\alpha \rightarrow 0$ is implied
in Eq.(56). With the definition of particle fields listed below$^{[2,3,8]}$ 
\begin{equation}
W_\mu ^{\pm }=\frac 1{\sqrt{2}}(A_\mu ^1\mp iA_\mu ^2)  \eqnum{57}
\end{equation}
\begin{equation}
\left( 
\begin{array}{c}
Z_\mu \\ 
A_\mu
\end{array}
\right) =\left( 
\begin{array}{c}
\cos \theta _w-\sin \theta _w \\ 
\sin \theta _w\text{ }\cos \theta _w
\end{array}
\right) \left( 
\begin{array}{c}
A_\mu ^3 \\ 
B_\mu
\end{array}
\right)  \eqnum{57}
\end{equation}
where $\theta _w$ is the Weinberg angle 
\begin{equation}
C^{\pm }=\frac 1{\sqrt{2}}\left( C^1\mp iC^2\right) \text{ , }\overline{C}%
^{\pm }=\frac 1{\sqrt{2}}(\overline{C}^1\mp i\overline{C}^2)  \eqnum{58}
\end{equation}
\begin{equation}
\left( 
\begin{array}{c}
C_z \\ 
C_\gamma
\end{array}
\right) =\left( 
\begin{array}{c}
\cos \theta _w-\sin \theta _w \\ 
\sin \theta _w\text{ }\cos \theta _w
\end{array}
\right) \left( 
\begin{array}{c}
C^3 \\ 
C^0
\end{array}
\right)  \eqnum{59}
\end{equation}
and 
\begin{equation}
\left( 
\begin{array}{c}
\overline{C}_z \\ 
\overline{C}_\gamma
\end{array}
\right) =\left( 
\begin{array}{c}
\cos \theta _w-\sin \theta _w \\ 
\sin \theta _w\text{ }\cos \theta _w
\end{array}
\right) \left( 
\begin{array}{c}
\overline{C}^3 \\ 
\overline{C}^0
\end{array}
\right)  \eqnum{60}
\end{equation}
the effective Lagrangian in Eq.(54) can be rewritten in the form 
\begin{equation}
{\cal L}_{eff}={\cal L}_g+{\cal L}_f+{\cal L}_H+{\cal L}_{gf}+{\cal L}_{gh} 
\eqnum{61}
\end{equation}
where 
\begin{eqnarray}
{\cal L}_g &=&-\frac 12W_{\mu \nu }^{+}W^{-\mu \nu }-\frac 14[Z^{\mu \nu
}Z_{\mu \nu }+A^{\mu \nu }A_{\mu \nu }]+M_w^2W_\mu ^{+}W^{-\mu }+\frac 12%
M_z^2Z^\mu Z_\mu  \nonumber \\
&&\ +ig[(W_{\mu \nu }^{+}W^{-\mu }-W_{\mu \nu }^{-}W^{+\mu })\left( \sin
\theta _wA^\nu +\cos \theta _wZ^\nu \right) +W_\mu ^{+}W_\nu ^{-}\left( \sin
\theta _wA^{\mu \nu }+\cos \theta _wZ^{\mu \nu }\right) ]  \nonumber \\
&&\ +g^2\{W_\mu ^{+}W_\nu ^{-}\left( \sin \theta _wA^\mu +\cos \theta
_wZ^\mu \right) (\sin \theta _wA^\nu +\cos \theta _wZ^\nu )-W_\mu
^{+}W^{-\mu }\left( \sin \theta _wA_\nu +\cos \theta _wZ_\nu \right) ^2 
\nonumber \\
&&\ +\frac 12[(W_\mu ^{+})^2\left( W_\nu ^{-}\right) ^2-(W_\mu ^{+}W^{-\mu
})^2]\}  \eqnum{62}
\end{eqnarray}
here 
\begin{eqnarray}
W_\mu ^{\pm } &=&(\partial _\mu W_\mu ^{\pm }-\partial _\nu W_\mu ^{\pm
}),Z_{\mu \nu }=\partial _\mu Z_\nu -\partial _\nu Z_\mu ,  \nonumber \\
A_{\mu \nu } &=&\partial _\mu A_\nu -\partial _\nu A_\mu  \eqnum{63}
\end{eqnarray}
and 
\begin{equation}
M_w=\frac 12gv,M_Z=M_w/\cos \theta _w  \eqnum{64}
\end{equation}
\begin{eqnarray}
{\cal L}_f &=&\overline{\nu }i\gamma ^\mu \frac 12(1-\gamma _5)\partial _\mu
\nu +\overline{l}(i\gamma ^\mu \partial _\mu -m_l)l+\frac g{\sqrt{2}}(j_\mu
^{-}W^{+\mu }+j_\mu ^{+}W^{-\mu })-ej_\mu ^{em}A^\mu  \nonumber \\
&&\ +\frac e{2\sin 2\theta _w}j_\mu ^0Z^\mu  \eqnum{65}
\end{eqnarray}
in which 
\begin{equation}
j_\mu ^{+}=\overline{l}\gamma _\mu \frac 12(1-\gamma _5)\nu =(j_\mu ^{-})^{+}
\eqnum{66}
\end{equation}
\begin{equation}
j_\mu ^{em}=\overline{l}\gamma _\mu l  \eqnum{67}
\end{equation}
\begin{equation}
j_\mu ^0=\overline{\nu }\gamma _\mu (1-\gamma _5)\nu -\overline{l}\gamma
_\mu (1-\gamma _5)l+4\sin ^2\theta _wj_\mu ^{em}  \eqnum{68}
\end{equation}
and 
\begin{equation}
m_l=\frac 1{\sqrt{2}}f_lv  \eqnum{69}
\end{equation}
\begin{equation}
{\cal L}_H=\frac 12\left( \partial ^\mu H\right) ^2-\frac 12m_H^2H^2+\frac g4%
(W^{+\mu }W_\mu ^{-}+\frac 1{2\cos \theta _w}Z^\mu Z_\mu )\left(
H^2+2vH\right) -\frac{f_l}{\sqrt{2}}\overline{l}lH-\lambda vH^3-\frac \lambda
4H^4  \eqnum{70}
\end{equation}
\begin{equation}
{\cal L}_{gf}=-\frac 1\alpha \partial ^\mu W_\mu ^{+}\partial ^\nu W_\nu
^{-}-\frac 1{2\alpha }(\partial ^\mu Z_\mu )^2-\frac 1{2\alpha }(\partial
^\mu A_\mu )^2  \eqnum{71}
\end{equation}
and 
\begin{eqnarray}
{\cal L}_{gh} &=&\overline{C}^{-}\Box C^{+}+\overline{C}^{+}\Box C^{-}+%
\overline{C}_z\Box C_z+\overline{C}_\gamma \Box C_\gamma  \nonumber \\
&&\ -ig\{(\partial ^\mu \overline{C}^{+}C^{-}-\partial ^\mu C^{-}C^{+})(\cos
\theta _wZ_\mu +\sin \theta _wA_\mu )  \nonumber \\
&&\ +(\partial ^\mu \overline{C}^{-}W_\mu ^{+}-\partial ^\mu \overline{C}%
^{+}W_\mu ^{-})\left( \cos \theta _wC_z+\sin \theta _wC_\gamma \right) 
\nonumber \\
&&\ +\left( \cos \theta _w\partial ^\mu \overline{C}_z+\sin \theta
_w\partial ^\mu \overline{C}_\gamma \right) (C^{+}W_\mu ^{-}-C^{-}W_\mu
^{+})\}  \eqnum{72}
\end{eqnarray}
The external source terms in Eq.(56) are defined by 
\begin{eqnarray}
J\cdot \Psi &=&J_\mu ^{-}W^{+\mu }+J_\mu ^{+}W^{-\mu }+J_\mu ^zZ^\mu +J_\mu
^\gamma A^\mu +JH+\overline{\xi }_ll  \nonumber \\
&&\ +\overline{l}\xi _l+\overline{\xi }_\nu \nu +\overline{\nu }\xi _\nu +%
\overline{\eta }^{+}C^{-}+\overline{C}^{+}\eta ^{-}++\overline{\eta }%
^{-}C^{+}+\overline{C}^{-}\eta ^{+}+\overline{\eta }_zC_z  \nonumber \\
&&\ +\overline{C}_z\eta _z+\overline{\eta }_\gamma C_\gamma +\overline{C}%
_\gamma \eta _\gamma  \eqnum{73}
\end{eqnarray}

The effective Lagrangian shown above is exactly the same as the one given in
the Landau gauge which was obtained in our preceding paper$^{[1]}$. In the
paper, the quantization of the electroweak theory without involving
Goldstone bosons was performed in the Lagrangian path-integral formalism by
the Faddeev-Popov's approach and/or the Lagrange multiplier method and the
quantum theory was given in the general $\alpha $-gauge . For the
quantization carried out in the Lagrangian path-integral formalism, it is
necessary to consider the SU(2)$\times $U(1) gauge-invariance of the
Lagrangian given in the unitary gauge and utilize the SU(2)$\times $U(1)
gauge transformations. Nevertheless, in the quantization performed in the
Hamiltonian path-integral formalism, as one has seen in this paper , we do
not need to consider any gauge transformation and the gauge- invariance of
the Lagrangian used. The same result obtained by the both quantizations
indicates that the electroweak theory without the Goldstone bosons surely
has the original SU(2)$\times $U(1) gauge symmetry. This feature of the
theory is natural because the theory can be written out from the ordinary $%
R_\alpha -$ gauge theory$^{[9,10]}$ by making the Higgs transformation to
the original Lagrangian and striking off the Goldstone fields from the
gauge-fixing terms and the ghost terms in the effective Lagrangian. This
procedure was justified in our preceding paper.

\section{\bf Acknowledgment}

The author wishes to thank professor Shi-Shu Wu for useful discussions. This
project was supported in part by National Natural Science Foundation of
China and The Research Fund for the Doctoral Program of Higher Education.

\section{\bf Appendix A: equations of Motion}

To help understanding the nature of the constraint conditions in Eqs.(10)
and (25), we briefly sketch the derivation of equations of motion for the
gauge fields. In order to get first order equations, the Lagrangian in
Eq.(2) will be recast in the first order form 
\begin{equation}
{\cal L}_g=\frac 12F^{a\mu \nu }F_{\mu \nu }^\alpha -\frac 12F^{a\mu \nu
}(\partial _\mu A_\nu ^a-\partial _\nu A_\mu ^a+g\epsilon ^{abc}A_\mu
^bA_\nu ^c)+\frac 14B^{\mu \nu }B_{\mu \nu }-\frac 12B^{\mu \nu }(\partial
_\mu B_\nu -\partial _\nu B_\mu )  \eqnum{A.1}
\end{equation}
where the variables $F_{\mu \nu }^\alpha ,A_\mu ^\alpha ,B_{\mu \nu }$ and $%
B_\mu $ are all treated as independent. Then, from the stationary condition
of the action $S_\lambda =\int d^4x{\cal L}_\lambda $ where ${\cal L}%
_\lambda $ was defined in Eqs.(1), (A.1), (3) and (4), it is not difficult
to derive the following equations 
\begin{equation}
\partial ^\nu F_{\mu \nu }^a=g\epsilon ^{abc}F_{\mu \nu }^bA^{c\nu }+\frac 12%
gL\gamma _\mu \tau ^aL+\frac 14g^2\phi ^{+}\phi A_\mu ^a+\frac 14gg^{\prime
}\phi ^{+}\tau ^a\phi B_\mu -\partial _\mu \lambda ^a  \eqnum{A.2}
\end{equation}
\begin{equation}
\partial ^\nu B_{\mu \nu }=-\frac 12g^{\prime }(L\gamma _\mu L+2\overline{l}%
_R\gamma _\mu l_R)+\frac 14g^{\prime 2}\phi ^{+}\phi B_\mu +\frac 14%
gg^{\prime }\phi ^{+}\tau ^a\phi A_\mu ^a-\partial _\mu \lambda ^0 
\eqnum{A.3}
\end{equation}
\begin{equation}
F_{\mu \nu }^a=\partial _\mu A_\nu ^a-\partial _\nu A_\mu ^a+g\epsilon
^{abc}A_\mu ^bA_\nu ^c  \eqnum{A.4}
\end{equation}
\begin{equation}
B_{\mu \nu }=\partial _\mu B_\nu -\partial _\nu B_\mu  \eqnum{A.5}
\end{equation}
and the Lorentz condition written in Eq.(10) as well as the equations for
the fermions and the Higgs particle which we do not list here for brevity.

Setting $\mu \left( \nu \right) =0$ and $k=1,2,3$, the above equations will
be separately written as follows 
\begin{equation}
\partial _0A_k^a=\partial _kA_0^a-F_{k0}^a+g\epsilon ^{abc}A_k^bA_0^c 
\eqnum{A.6}
\end{equation}
\begin{equation}
\partial _0B_k=\partial _kB_0-B_{k0}  \eqnum{A.7}
\end{equation}
\begin{equation}
\partial _0F_{k0}^a=\partial ^lF_{lk}^a+g\epsilon
^{abc}(F_{k0}^bA_0^c+F_{kl}^bA^{cl})+\frac 12g\overline{L}\gamma _k\tau ^aL+%
\frac 14g^2\phi ^{+}\phi A_k^a+\frac 14gg^{\prime }\phi ^{+}\tau ^a\phi
B_k-\partial _k\lambda ^a  \eqnum{A.8}
\end{equation}
\begin{equation}
\partial _0B_{k0}=\partial ^lB_{lk}-\frac 12g^{\prime }\left( \overline{L}%
\gamma _kL+2\overline{l}_R\gamma _kl_R\right) +\frac 14gg^{\prime }\phi
^{+}\tau ^a\phi A_k^a+\frac 14g^{\prime 2}\phi ^{+}\phi B_k-\partial
_k\lambda ^0  \eqnum{A.9}
\end{equation}
\begin{equation}
F_{kl}^a=\partial _kA_l^a-\partial _lA_k^a+g\epsilon ^{abc}A_k^bA_l^c 
\eqnum{A.10}
\end{equation}
\begin{equation}
B_{kl}=\partial _kB_l-\partial _lB_k  \eqnum{A.11}
\end{equation}
\begin{equation}
\partial ^kF_{k0}^a=g\epsilon ^{abc}F_{k0}^bA^{ck}-\frac 12g\overline{L}%
\gamma _0\tau ^aL-\frac 14g^2\phi ^{+}\phi A_0^a-\frac 14gg^{\prime }\phi
^{+}\tau ^a\phi B_0+\partial _0\lambda ^a  \eqnum{A.12}
\end{equation}
\begin{equation}
\partial ^kB_{k0}=\frac 12g^{\prime }\left( \overline{L}\gamma _0L+2%
\overline{l}_R\gamma _0l_R\right) -\frac 14gg^{\prime }\phi ^{+}\tau ^a\phi
A_0^a-\frac 14g^{\prime 2}\phi ^{+}\phi B_0+\partial _0\lambda ^0 
\eqnum{A.13}
\end{equation}
As we see, in Eqs.(A.6)-(A.9) there are the time-derivatives of the field
variables $A_k^a,B_k,F_{k0}^a=E_k^a$ and $B_{k0}=E_k^0$, whereas in
Eqs.(A.10)-(A.13), there are no such derivatives. Therefore, Eqs.(A.6)-(A.9)
act as the equations of motion, while, Eqs.(A.10)-(A.13) can only be
identified with the constraint equations. With the definitions given in
Eqs.(12) and (21-(24), we see that Eqs.(A.12) and (A.13) are just combined
to give the constraint equation written in Eq.(25).

\section{\bf Appendix B:The Delta-Functional Representation of Constraint
Equations}

In this appendix, we take an example to prove the relation shown in Eq.(38).
Suppose we have two equations 
\begin{equation}
\varphi _1(x,y)=0  \eqnum{B.1}
\end{equation}
\begin{equation}
\varphi _2(x,y)=0  \eqnum{B.2}
\end{equation}
whose solutions are assumed to be ($x_s,y_s)$. Let us evaluate the integral 
\begin{equation}
I=\int dxdyf(x,y)\delta [\varphi _1(x,y)]\delta [\varphi _2(x,y)] 
\eqnum{B.3}
\end{equation}
where $f(x,y)$ is an arbitrary integrable function. This integral may be
easily calculated by making the change of the integration variables 
\begin{equation}
\varphi _1(x,y)=u_1\text{ , }\varphi _2(x,y)=u_2  \eqnum{B.4}
\end{equation}
Correspondingly, the integration measure will be changed to 
\begin{equation}
dxdy=\det \left( \frac{\partial (\varphi _1,\varphi _2)}{\partial (x,y)}%
\right) ^{-1}du_1du_2  \eqnum{B.5}
\end{equation}
Substituting Eqs.(B.4) and (B.5) in Eq.(B.3), we have 
\begin{eqnarray}
I &=&\int du_1du_2\det \left( \frac{\partial (\varphi _1,\varphi _2)}{%
\partial (x,y)}\right) ^{-1}\delta \left( u_1\right) \delta \left(
u_2\right) f(x(u_1,u_2),y(u_1,u_2))  \nonumber \\
&=&\det \left( \frac{\partial (\varphi _1,\varphi _2)}{\partial (x,y)}%
\right) ^{-1}f(x(u_1,u_2),y(u_1,u_2))_{|u_1=u_2=0}  \eqnum{B.6}
\end{eqnarray}
Noticing 
\begin{equation}
x\left( u_1,u_2\right) _{|u_1=u_2=0}=x_s\text{ , }y\left( u_1,u_2\right)
_{\mid u_1=u_2=0}=y_s  \eqnum{B.7}
\end{equation}
we can write 
\begin{equation}
I=\det \left( \frac{\partial (\varphi _1,\varphi _2)}{\partial (x,y)}\right)
_{|x=x_s,y=y_s}^{-1}f(x_s,y_s)=\int dxdyf(x,y)\det \left( \frac{\partial
(\varphi _1,\varphi _2)}{\partial (x,y)}\right) ^{-1}\delta \left(
x-x_s\right) \delta (y-y_s)  \eqnum{B.8}
\end{equation}
In comparison of this expression with that denoted in Eq.(B.3), it is clear
to see that 
\begin{equation}
\delta \left( x-x_s\right) \delta (y-y_s)=\det \left( \frac{\partial
(\varphi _1,\varphi _2)}{\partial (x,y)}\right) \delta (\varphi _1)\delta
(\varphi _2)  \eqnum{B.9}
\end{equation}
For delta-functionals, certainly, we have the same relation, just as shown
in Eq.(38)

\section{References}

[1] J. C. Su, Alternative Formulation of The Electroweak Theory, hep. th
/00100180.

[2] S. Weinberg, Phys. Rev. Lett. 19, 1264 (1967).

[3] A. Salam, Elementary Particle Theory, ed. by N. Svartholm, Almqvist and
Wiksell, Stockholm, 367 (1968).

[4] L. D. Faddeev and V. N. Popov, Phys. Lett. B25, 29 (1967).

[5] L. D. Faddeev, Theor. Math. Phys. 1, 1 (1970).

[6] L. D. Faddeev and A. A. Slavnov, Gauge Fields: Introduction to Quantum
Theory, The Benjamin Commings

\ Publishing Company Inc. (1980).

[7] P. Senjanovic, Ann. Phys. (N.Y) 100, 227 (1976).

[8] M. E. Deskin and D. V. Schroeder, An Introduction to Quantum Field
Theory, Addison-Wesley Publishing

Company, (1995).

[9] K. Fujikawa, B. W. Lee and A. I. Sanda, Phys.Rev. D6, 2923 (1972).

[10] G. 't Hooft and M. J. G. Veltman, Nucl. Phys. B35, 167 (1971).

\end{document}